\documentclass[%
reprint,
twocolumn,
showpacs,preprintnumbers,
amsmath,amssymb,
aps,pre, 
nofootinbib,
longbibliography, 
floatfix,
]{revtex4-2}

\usepackage{units}

\usepackage{upgreek}
\usepackage{textcomp}
\usepackage{enumitem}  
\usepackage{graphicx}
\usepackage{epstopdf}
\usepackage{dcolumn}
\usepackage{bm}
\usepackage{hyperref}
\usepackage{booktabs}

\usepackage{xcolor}
\definecolor{mylinkcolor}{rgb}{0.0,0.0,0.66}
\hypersetup{colorlinks=true,linkcolor=mylinkcolor,urlcolor=mylinkcolor,citecolor=mylinkcolor,filecolor=mylinkcolor,breaklinks=true,linktocpage=true,linktoc=all}

\usepackage[normalem]{ulem}
\usepackage{color,soul} 
\usepackage{lipsum}  
\usepackage[export]{adjustbox}

\newcommand{\ie}{\textit{i.e.,}\ }

\newcommand{\sdist}{\kern 0.20em}
\renewcommand{\eqref}[1]{Eq.\sdist(\ref{#1})}

\mathchardef\mhyphen="2D

\usepackage{color}

\emergencystretch=\maxdimen
\begin{document}


\title{Producing entangled photon pairs and quantum squeezed states  in plasmas}
\author{Kenan Qu}
\affiliation{Department of Astrophysical Sciences, Princeton University,  Princeton, New Jersey 08544, USA \looseness=-1 }  
\author{Nathaniel J. Fisch}
\affiliation{Department of Astrophysical Sciences, Princeton University,  Princeton, New Jersey 08544, USA \looseness=-1 }

\date{\today}

\begin{abstract}
Plasma is capable of mediating the conversion of two pump photons into two different photons through a relativistic four-wave mixing nonlinearity. Spontaneously created photon pairs are emitted at symmetric angles with respect to the colinear pump direction, and the emission rate is largest if they have identical frequency. Thus, two orthogonally polarized pumps can produce polarization-entangled photon pairs through a mm-long homogeneous plasma. The noise from Raman scattering can be avoided if the pump detuning differs from twice the plasma frequency. On the other hand, pump detuning exactly equal to twice the plasma frequency can significantly enhance the interaction rate, which allows for the production of strong two-mode squeezed states. Remarkably, the amplified noise from Raman scattering are correlated and hence can be suppressed in one of the output quadratures, thereby maintaining the squeezing magnitude.
\end{abstract}


\maketitle

\section{Introduction}

Quantum entangled photon pairs and quantum squeezed states are two types of the most crucial resources for quantum information science. Entangled photon pairs have non-local correlations to enable a variety of applications like quantum computing and communication. Quantum squeezed states exhibit a lower noise level in one quadrature than the vacuum state, offering a significant advantage in high-precision measurements. Notably, its application in advanced-LIGO detectors~\cite{np_LIGO_2013,PRX_LIGO_2023} has demonstrably boosted detection rates by over 60\%. 
However, the advantages of utilizing quantum non-classical light are constrained by low photon flux and narrow bandwidth. This limitation results in low frame rates, typically a fraction of a hertz, in quantum imaging~\cite{Moodley_23} and quantum spectroscopy~\cite{Mukamel_2020} experiments, due to the restricted photon generation rates. Similarly, the SU(1,1) interferometer~\cite{SU11_Yurke1986, Ou_APL2020} has yet to surpass the conventional SU(2) interferometer due to limited squeezing performance.

Production of entangled photons and squeezed light typically uses spontaneous parametric down-conversion in nonlinear crystals which, in a classical description~\cite{Boyd_NO_book}, arises from the anharmonic potential of the crystal electrons in a strong driving laser field. 
Efforts to enhance the photon flux and bandwidth of non-classical light generation fall into two categories. First, the nonlinear optics community focuses on optimizing conventional nonlinear crystals through techniques like periodic poling~\cite{Franken_RMP1963,Canagasabey_09}. This method effectively increases photon emission rates by achieving quasi-phase matching and minimizing phase drift.
Second, researchers have explored alternative systems with higher nonlinear optical susceptibilities, including optical fibers~\cite{Lin_06fiber, Garay_Palmett:23}, silicon waveguides~\cite{Lin_06silicon, semiconductor_2004, Chopin_waveguide}, superconducting Josephson junctions~\cite{Peugeot_JJ2021}, cold atoms~\cite{Riebe_2004, Ghosh_2021, Craddock_PRApp2024}, and optomechanical systems~\cite{Purdy_2013, Qu_NJP2014, Qu_PRA2015, Magrini_2022, PRXQMicro_2021}.  However, all these approaches employ weak laser fields, fundamentally restricting the output photon flux.
Recent advancements in attosecond physics have spurred investigations into high harmonic generation~\cite{nc_HHG2017, np_Lewenstein2021, np_HHG2023, PRL_Tsatrafyllis2019}  and its potential for nonclassical light production using laser intensities of $\unit[10^{12-14}]{Wcm^{-2}}$~\cite{sc_HHG2016, nc_HHG2017, PRL_Tsatrafyllis2019, nc_HHG2020}. While the non-classical nature of high harmonic photons holds promise for testing quantum theory and studying electron interactions with strong quantum light~\cite{np_Even2023}, their practical applications in quantum optics remain unclear.

Further increasing laser intensity, however, causes thermal damage to conventional nonlinear materials. Plasmas, on the other hand, can maintain optical properties above the ionization laser intensity. This high thermal damage threshold positions plasma as a potential candidate for delivering the next generation of high-intensity laser sources~\cite{Shvets_1998, malkin99, Cheng2005, Ren_np2007, Ren_PoP2008, Qu_PRL2017}. 
Additionally, plasmas exhibit strong nonlinearity~\cite{Joshi_1990} at high intensities, allowing rapid amplification~\cite{malkin99, PRL_Cheng2005, Ren_np2007} and storage~\cite{DODIN200283, Dodin_PRL2002, Lehmann_PRL2016, Lehmann_PRE2018, Lehmann2019, Lehmann2023} of light pulses, and merging laser energy of multiple kJ~\cite{np_Kirkwood2018}.  Importantly, the plasma nonlinearity scales across a broad range of frequencies, enabling manipulation from microwaves to X-rays. 

This paper investigates the use of the relativistic four-wave mixing (FWM) nonlinearity of plasmas to produce quantum entangled photon pairs and squeezed states. Several plasma experiments since the 1980s have demonstrated degenerate FWM~\cite{Steel_79, Federici_86, Domier1993, Gupta1998, Lee_PRE2007}  using relatively low laser powers. However, these approaches employed a Brillouin grating generated through the laser ponderomotive force, which introduces classical noise and hence is not suitable for directly producing quantum light. 

Recently, however, for the purposes of laser upconversion, all-optical parametric processes have been proposed at high power in under-dense plasma, in which the plasma is used for coupling electromagnetic pulses without affecting the resonance condition.  In this way, the plasma can efficiently mediate the conversion of near-optical laser pulses to high-energy X-rays in a cascaded manner~\cite{Malkin_pre2020, Malkin_pre2020_2, Malkin_pre2022, Malkin_pre2023, Griffith2021}.  In particular, the relativistic FWM in plasma was proposed for converting two pump photons into two output photons at different frequencies in under-dense plasma. The relativistic FWM process uses a $\chi^{(3)}$ nonlinearity which couples four electromagnetic waves through anharmonic electron motion caused by the relativistic effects in the strong laser field. In fact, the similar FWM process in optical fibers~\cite{Lin_06fiber, Garay_Palmett:23} has been established as a key technique for generating broadband entangled photon pairs. However, in plasma, the tolerance to extreme laser intensities enables ultrahigh 
FWM interaction rate with a cm-scale growth length. Crucially, the nonlinear plasma dispersion relation can be utilized to create phase matched interaction through arranging the laser wavevectors and plasma density. Entangled photon pairs can then be produced in a sub-mm-long plasma and strong squeezing can be obtained in a longer plasma. 

The all-optical possibilities for FWM are crucial in realizing entangled photons that survive plasma noise.   The faster plasma Raman scattering process, which is a lower order parametric process, is subject to classical noise effects. While the thermal bath for unseeded FWM is the vacuum fluctuation, the thermal bath for spontaneous Raman scattering (SRS) is the random fluctuation of plasma density or thermal phonons.  In fact, the fast growth rate of SRS is a significant challenge for many plasma photonic applications~\cite{Malkin2000, MinSup_np2023}, particularly in high-density plasmas.

This paper demonstrates two methods of suppressing the SRS noise. The first method takes advantage of the discrete emission spectrum of SRS, i.e., it emits to only the Stokes and anti-Stokes side bands of the pump wave. The special phase matching condition, reported in Refs.~\cite{Malkin_pre2020, Malkin_pre2020_2}, offers a route to tailor the FWM emission frequency by tuning the pump frequencies and the plasma density. Thus, SRS can be effectively suppressed by detuning the pumps away from the side bands of the output modes. Because detuning from plasma resonance reduces the FWM growth rate, this approach is particularly effective only for generating single-photon level output. The second method exploits the correlation and cancellation of the SRS noise in both output modes.
Specifically, when two pumps have equal amplitudes and their beat frequency matches twice the plasma frequency, the FWM growth rate is maximized, and the system functions as a two-mode squeezer and a phase-sensitive amplifier. While plasma wave phonons are created from scattering of the higher frequency pump, they are simultaneously annihilated by interacting with the lower frequency pump, thereby maintaining a fixed amplitude. Furthermore, the amplified plasma waves couple exclusively to one quadrature of the optical state, suppressing the SRS noise in the squeezed quadrature.

The paper is organized as follows: In Sec.~II, we analyze two types of interactions between intense laser pulses and plasmas. Through quantization of the plasma wave, we obtain the interaction Hamiltonian of the laser-plasma system. In Sec.~III, we investigate the production of polarization entangled photon pairs using the relativistic FWM nonlinearity of plasmas. The condition for suppressing SRS is analyzed and the logarithmic negativity is obtained. In Sec.~IV, we demonstrate how a quantum two-mode squeezed state can be produced using two colinear pumps with a frequency detuning equal to twice the plasma frequency. The solution to the quantum Langevin equations yields the squeezing magnitude and its degradation due to thermal phonons. We show that balanced pump strength can effectively reduce the noise in one quadrature. In Sec.~V, we present our conclusions and discussions.

\section{Laser plasma interactions and Hamiltonian}

Fully ionized plasmas, consisting of both electrons and ions (or positrons), can mediate laser interactions through a variety of processes. For the purpose of creating entangled photon pairs and quantum squeezed light, we focus on parametric processes that have fast growth rates. For this purpose, we analyze the motion of electrons in a laser field which creates a polarization current  $\bm{J}$ that drives the laser field $ \bm{E}$ through the wave equation 
\begin{equation}\label{eq:E}
	(\frac{\partial^2}{\partial t^2} - c^2\frac{\partial^2}{\partial z^2})  \bm{E}  = 
	\frac{1}{\varepsilon_0} \frac{\partial}{\partial t} \bm{J}, 
\end{equation}
where $c$ is the speed of light in vacuum and $\varepsilon_0$ is the vacuum permittivity. In the simplest ``fluid'' model of plasmas, we neglect electron kinetic effects. The acceleration of the polarization current can then be written as
\begin{equation}
	\frac{\partial}{\partial t} \bm{J} = \frac{e^2 n_e}{m_e\gamma} \bm{E},
\end{equation}
where $e$ is the natural charge, $n_e$ is the electron number density, $m_e$ is the electron rest mass, and $\gamma$ is the relativistic Lorentz factor. 

The explicit inclusion of $\gamma$ reveals the first laser plasma interaction: If the laser field is sufficiently strong to drive the electron to relativistic speeds, the electron mass increases by a factor $\gamma$ near the laser anti-nodes when it reaches its maximum kinetic energy. Assuming the electrons start from rest, the value of $\gamma$ can be obtained using the conservation of canonical momentum, \ie $\gamma(t) \approx \sqrt{1+\alpha^2(t)}$, where we introduced the parameter $\bm\alpha = e\bm{A}/(m_ec^2)$ and its amplitude $\alpha$ to denote the normalized laser amplitude.~\footnote{It is usually denoted as $a$ (or $a_0$ for its peak value) in the plasma community, but we reserve symbol $a$ for quantum fields in this work.} Here, $\bm{A}$ is the laser vector potential. We can thus expand $1/\gamma \approx 1- \alpha^2/2$ for $\alpha<1$. In a physical picture, the relativistic effect causes an anharmonic electron oscillation to lead to scatterings at different wavelengths. 

The second type of interaction is associated with the fluctuating plasma density $n_e$. The displacement of an electron from its stable position causes a electrostatic restoring force. If all the plasma electrons are driven by an external field, such as the laser ponderomotive force, the electrostatic restoring force is amplified to drive the electrons into a longitudinal oscillation, called a plasma Langmuir wave. The plasma could thus exhibit a spatial density modulation which itself oscillates at the plasma frequency $\omega_p = \sqrt{e^2 n_e/(m_e\varepsilon_0)}$. The electron density ripple, functioning as a Bragg grating, scatters the incoming laser field. At the same time, the oscillation motion Doppler shifts the laser frequency to cause a Stokes side band and an anti-Stokes side band with a frequency detuning equal to $\pm \omega_p$. The frequency-detuned scattered light beats with the drive laser and reinforces the plasma oscillation. The positive feedback results in an instability, called Raman scattering. This scattering is further referred to as spontaneous Raman scattering and stimulated Raman scattering, depending on whether a seed is used to initiate the instability. Here, we use the the acronym SRS for both scatterings. 

\begin{figure}[thp]
	\centering
	\includegraphics[width=0.75\linewidth]{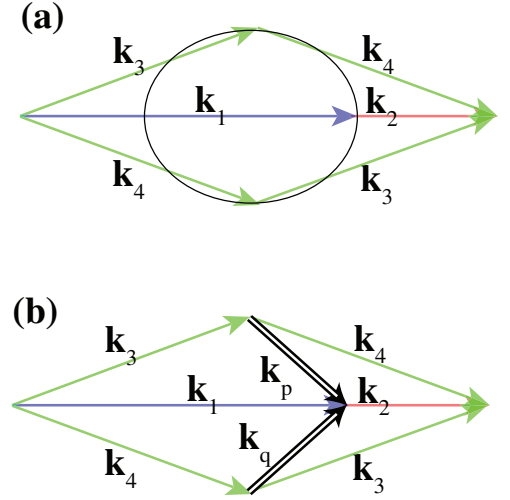}
	\caption{(a) Wavevector relations of FWM. Pump photons ($\bm{k}_1$ and $\bm{k}_2$)  are directly converted into emitted photons ($\bm{k}_3$ and $\bm{k}_4$) if they satisfy the Manley-Rowe relation. (b) Wavevector relation of SRS involving four electromagnetic waves and two plasma waves. Pump photons are converted into emitted photons through plasma waves  ($\bm{k}_p$ and $\bm{k}_q$) if $\omega_{1,2}\pm \omega_{3,4} = \omega_p$ is also satisfied. } 
	\label{fig:diagram_FS}
\end{figure}

Writing $n_e = \bar{n}_e(1+\delta n)$, the wave equation (\ref{eq:E}) can be transformed into
\begin{equation}
	(\frac{\partial^2}{\partial t^2} - c^2\frac{\partial^2}{\partial z^2}) \bm\alpha = (1+\delta n - \frac{\alpha^2}{2}) \bm\alpha. 
\end{equation}
The first-order Taylor expansion of $1/\gamma$ shows a FWM coupling if the Manley-Rowe relations are satisfied, \ie $\omega_1+\omega_2 = \omega_3 + \omega_4$ and $\bm{k}_1 + \bm{k}_2 = \bm{k}_3 + \bm{k}_4$. Taking into account  the plasma dispersion relation $\omega^2 = c^2k^2 + \omega_p^2$, the allowed wavevectors trace an ellipsoid as represented in 2D in Fig.~\ref{fig:diagram_FS}(a). 

The laser wave equation is accompanied by the plasma wave equation 
\begin{equation}
	(\frac{\partial^2}{\partial t^2} + \omega_p^2)  \delta n = \frac{c^2}{2} \nabla^2 \alpha^2, 
\end{equation}
which describes the driven motion of plasma density modulation by the laser ponderomotive force. Frequency matching condition shows that SRS takes place if two electromagnetic modes are detuned by the plasma frequency $\omega_p$, which is the eigenfrequency of plasma oscillations and is independent of wavevector in a cold plasma. We illustrate the phase matching condition in Fig.~\ref{fig:diagram_FS}(b).

To quantize the electromagnetic wave and the plasma wave (variables $\alpha$ and $\delta n$) and to obtain the interaction Hamiltonian, we expand the wave equations to the first order using the slowly varying envelope approximation and the plasma dispersion relation. 
Including the plasma waves, the modes of interest can all be expanded as
\begin{align}
	\alpha &= \sum_{i=1}^4 \alpha_i e^{i\bm{k}_i\bm{r}-i\omega_i t} + c.c. \\
	\delta n &= \delta n_p e^{i\bm{k}_p\cdot \bm{r} - i\omega_pt} + \delta n_q  e^{i\bm{k}_q\cdot \bm{r} - i\omega_pt}  + c.c.. 
\end{align}
Now we consider two pump modes with frequencies $\omega_{1,2}$ and two output modes with frequencies $\omega_{3,4}$. Their wavevector relations are sketched in Fig.~\ref{fig:diagram_FS}. Then, we obtain the equations of motion after neglecting the fast rotating terms
\begin{align}
	(\frac{\partial}{\partial t} - \bm{v}_3\cdot \nabla) \alpha_3 &= \frac{i\omega_p^2}{\omega_3} (\alpha_1\alpha_2 \alpha_4^* - \alpha_1 \delta n_p^* - \alpha_2 \delta n_q), \\
	(\frac{\partial}{\partial t} - \bm{v}_4\cdot \nabla) \alpha_4 &= \frac{i\omega_p^2}{\omega_4} (\alpha_1\alpha_2 \alpha_3^* - \alpha_1 \delta n_q^* - \alpha_2 \delta n_p), \\
	\frac{\partial}{\partial t} \delta n_p &= \frac{i c^2}{\omega_p} \nabla^2 (\alpha_1\alpha_3^* + \alpha_2^*\alpha_4) ,  \\
	\frac{\partial}{\partial t} \delta n_q &= \frac{i c^2}{\omega_p} \nabla^2 (\alpha_1\alpha_4^* + \alpha_2^*\alpha_3). 
\end{align}
The pump fields $\alpha_{1,2}$ have near relativistic intensities ($\unit[10^{16}-10^{17}]{Wcm^{-2}}$), so they can be treated classically. 
The variables to be quantized are $\alpha_{3,4}$ and $\delta n_{p,q}$. 
The field $\alpha_{3,4}$ can be quantized using the standard procedure~\cite{Agarwal_2012} $\bm{A} = \sum_{\bm{k},s} \sqrt\frac{\hbar}{2\omega_{\bm{k},s}V\varepsilon_0}  \hat{a}_{\bm{k},s} \hat{\sigma}_{\bm{k},s} e^{i\bm{k}\cdot\bm{r} - i\omega t} + h.c.$ where $\hat{a}_{\bm{k},s}$ is the annihilation operator, $V$ is the normalization volume, and $\hat\sigma_i$ is the polarization vector. 
The plasma wave can be treated as a phonon which interacts with the laser by absorbing or emitting a photon. 
To quantize the phonon mode of the plasma wave, we use the fact that the interaction Hamiltonian has the form 
\begin{gather} \label{eq:hamil}
	H_{int} = H_{FWM} + H_{SRS}, \\
	H_{FWM} = \hbar\Gamma \alpha_1\alpha_2 \hat{a}_3^\dag \hat{a}_4^\dag + h.c, \\
	 H_{SRS} = \hbar g   [\alpha_1^* (\hat{a}_3 \hat p + \hat{a}_4 \hat q) + \alpha_2^* (\hat{a}_3 \hat q^\dag + \hat{a}_4 \hat p^\dag)]+ h.c.,
\end{gather}
where $\Gamma = \frac{\omega_p^2}{2\sqrt{\omega_3\omega_4}} \sum_{i,j,k,l=1,2,3,4}(\hat\sigma_i \cdot \hat\sigma_j) (\hat\sigma_k \cdot \hat\sigma_l)$. 
It thus leads to the relations $|\delta n_p|^2 \leftrightarrow \frac{\hbar e^2 k_p^2}{2V\epsilon_0 \omega_p^3m_e^2c^2} \hat p^\dag \hat p$ and $ g  = \frac{ck_p}{2}\sqrt\frac{\omega_p}{\omega_3}$. Because $k_p = k_q$ and $\omega_3=\omega_4$, the normalization is the same for mode $\hat q$. A different normalization would change the zero point fluctuation energy but would not alter the key result because the plasma wave amplitude would not grow exponentially  as we will show.

\section{Producing polarization entangled photon pairs} \label{sec:FWM}

The interaction Hamiltonian (\ref{eq:hamil}) describes two processes of creating photon pairs, including using FWM and using phonon-mediated SRS. FWM is a parametric process and the electrons do not change their states after absorbing a pair of pump photons and emitting another pair of output photons. Thus, the combined property of the output photons, including their frequencies, wavevectors, polarization, and emission angles, will be identical to absorbed pump photon pair. The output states under the FWM interaction can be expressed as $
|\Psi\rangle = \exp[-(\frac{iz}{\hbar c}) H_{FWM} ] |0,0\rangle  
= \cosh^{-1}r \sum_{n=0}^\infty e^{in\varphi}\tanh^n r |n,n\rangle$.
Here, the quantum squeezing parameter and phase is related to the interaction time via $ \Gamma \alpha_1\alpha_2 z/c = r e^{i\varphi}$. The two output modes have an identical photon number and are quantum correlated. For short plasma or weak pump fields $r \ll 1$, the output field becomes a single photon pair $|1,1\rangle$. 
Variation of any parameter of one of the output photon will be correlated with the other photon. This process offers a mechanism to produce entangled photon pairs. 

On the other hand, SRS is an instability which can grow from scattering and amplifying a plasma wave. Because a plasma Langmuir wave is an eigenmode of the plasma medium, it could exist due to thermal effect and hence has a finite phonon number at finite temperatures. SRS of the pump wave is a fast process that creates photons that are not correlated other photons and reduces the quantum purity of the output states. 
Considering degenerate frequency for the entangled photon pairs $\omega_3=\omega_4$, the FWM growth rate is $\alpha_1\alpha_2\Gamma = \alpha_1\alpha_2 3\omega_p^2/(2\omega_3)$, but the SRS growth rate is $\alpha_{1,2} g \approx \alpha_{1,2} \omega_p^{3/2}/(2\omega_3)$. For plasma frequency $\omega_p\approx \omega_3/10$ and moderately intense laser $\alpha_{1,2}\approx 0.1$, the SRS growth rate is larger than FWM by a factor of $10$. 
Moreover, SRS of the pump could not be suppressed using techniques for plasma Raman amplifiers, such as a plasma density gradient, because fluctuations of the plasma density could broaden the scattering spectrum and shadow the entangled photon pairs.

Therefore, the noise from SRS needs to be reduced by choosing the pump parameters such that the output photon frequency is detuned sufficiently far from the Stokes or anti-Stokes side band of each pump pulse. Such an arrangement spectral isolates the pump scattering from FWM and from SRS. The amount frequency detuning required depends on the linewidth of plasma resonance. For cold plasmas with Debye length longer than the plasma wavelength, the plasma density fluctuation spectrum is dominated~\cite{ICHIMARU196278, Salpeter_PR1960} by the collective dynamics and has a  Lorentz shape. Its linewidth comes from Landau damping which is contributed by those few electrons in the tail of the Maxwell distribution whose velocity equals the phase velocity of the plasma oscillation. For plasma temperature of $10-100$eV and $k_p\sim \omega_1/(10c)$, the Landau damping is negligible and hence SRS is negligible as long as the detuning exceeds the sum of plasma frequency and the laser linewidth. 

Although plasma waves are not resonantly excited in this regime, they nevertheless influence the FWM growth rate by inducing plasma oscillations (albeit not at the plasma frequency). It is pointed out in Ref.~\cite{Malkin_pre2020} that the FWM growth rate is to be changed to
\begin{align}
	\Gamma_F &= \frac{\omega_p^2}{2\sqrt{\omega_3\omega_4}} (f_{1,2} + f_{1,-3} + f_{1,-4} ), \label{eq:F} \\
	f_{i,j} &= \left[ \frac{c^2(\bm{k}_i+\bm{k}_j)^2}{(\omega_i+\omega_j)^2 -\omega_p^2} -1 \right] (\hat\sigma_i \cdot \hat\sigma_j) (\hat\sigma_k \cdot \hat\sigma_l), \label{eq:f}
\end{align}
where we use the notation $\omega_{-i}=-\omega_i$ and $\bm{k}_{-i} = -\bm{k}_i$. As two mode detuning $\omega_i - \omega_j$ approaches the plasma frequency $\omega_p$, their beat drives plasma density oscillation which affects the FWM interaction rate. 

The polarization and wavevector of the emitted photon pairs are determined by the  FWM interaction rate $\Gamma_F$. 
For two pumps with identical polarization, all three terms in Eq.~(\ref{eq:F}) contribute to $\Gamma_F$, and both output photons have the same polarization. 
But if the two pumps have orthogonal polarization, only $f_{1,-3}$ and $f_{1,-4}$ are nonzero, and they correspond to different quantum paths of photon generation: $f_{1,-3}$ is proportional to the probability that modes 1 and 3, and modes 2 and 4, have the same polarization; $f_{1,-4}$ is proportional to the probability that modes 1 and 4, and modes 2 and 3, have the same polarization. 



\begin{figure}[thp]
	\centering
	\includegraphics[width=0.9\linewidth]{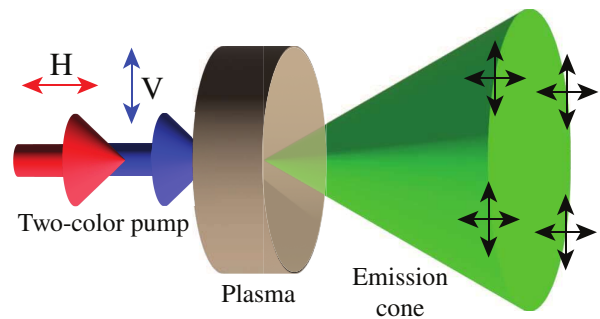}
	\caption{Two-color pump with orthogonal polarization creates polarization entangled photon pairs at symmetric angles. The shades represent the most probable emission angles for given pump polarization.. } 
	\label{diagram}
\end{figure}

\subsection{Pumps with orthogonal polarization}
We first consider using two colinear pump pulses with orthogonal polarization, as sketched in Fig.~\ref{diagram}. Their interaction in plasma produces photon pairs if their wavevectors satisfy the Manley-Rowe relations
\begin{gather}
	\omega_1+\omega_2 = \omega_3 + \omega_4, \\
	\omega_3 = \sqrt{c^2(\frac{k_1+k_2}{2} + q)^2 + c^2k_\perp^2 + \omega_p^2},  \\
	\omega_4 =  \sqrt{c^2(\frac{k_1+k_2}{2} - q)^2 + c^2k_\perp^2 + \omega_p^2},
\end{gather}
where $q=(k_{3\|} - k_{4\|})/2$. 
The possible combination of wavevectors traces an ellipse, as plotted in Fig.~\ref{fig:diagram_FS}(a). 
Photon pairs of the same frequency are emitted at angle $\alpha$ such that $k_3=k_4$
\begin{equation}\label{eq:alpha}
	\cos\alpha = \frac{c(k_1+k_2)}{\sqrt{(\omega_1+\omega_2)^2 - 4\omega_p^2}} .
\end{equation}

\begin{figure}[thp]
	\centering
	\includegraphics[height=7cm]{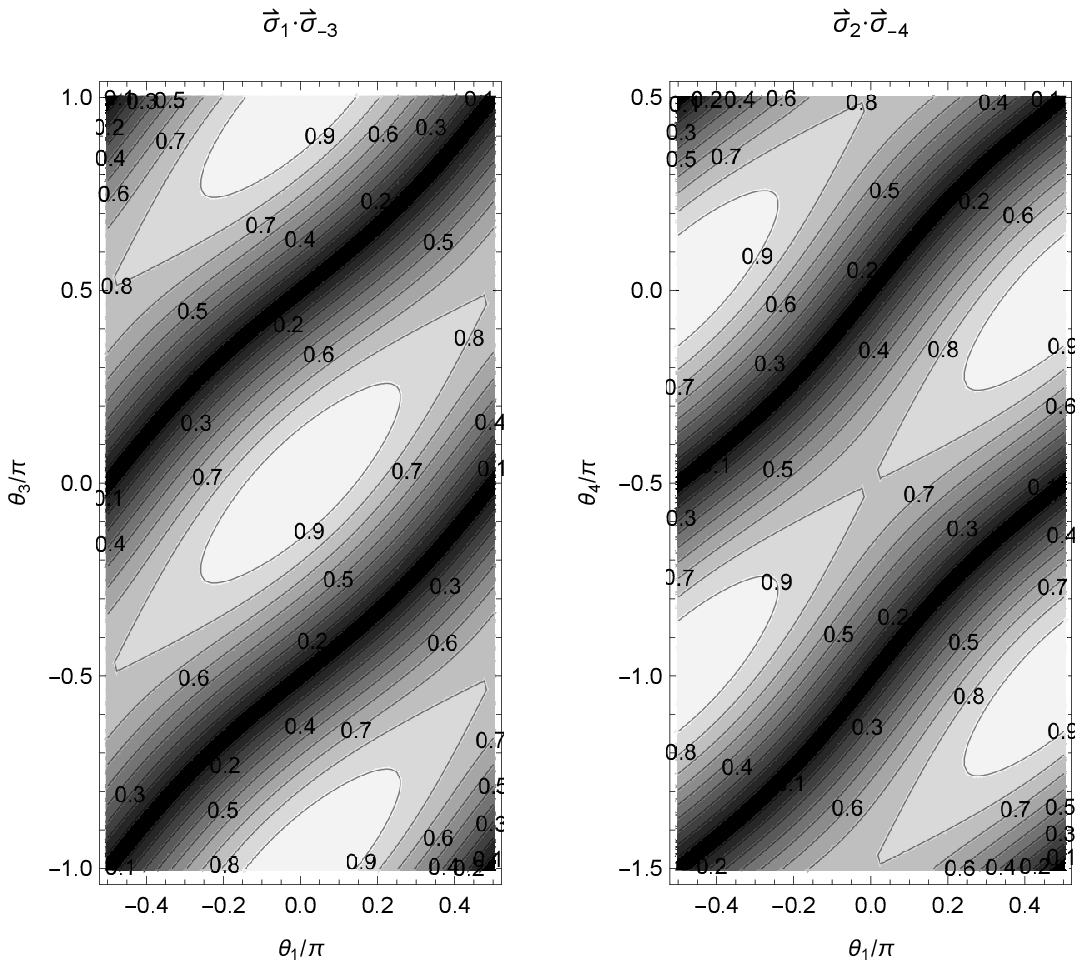} \quad
	\includegraphics[height=7cm]{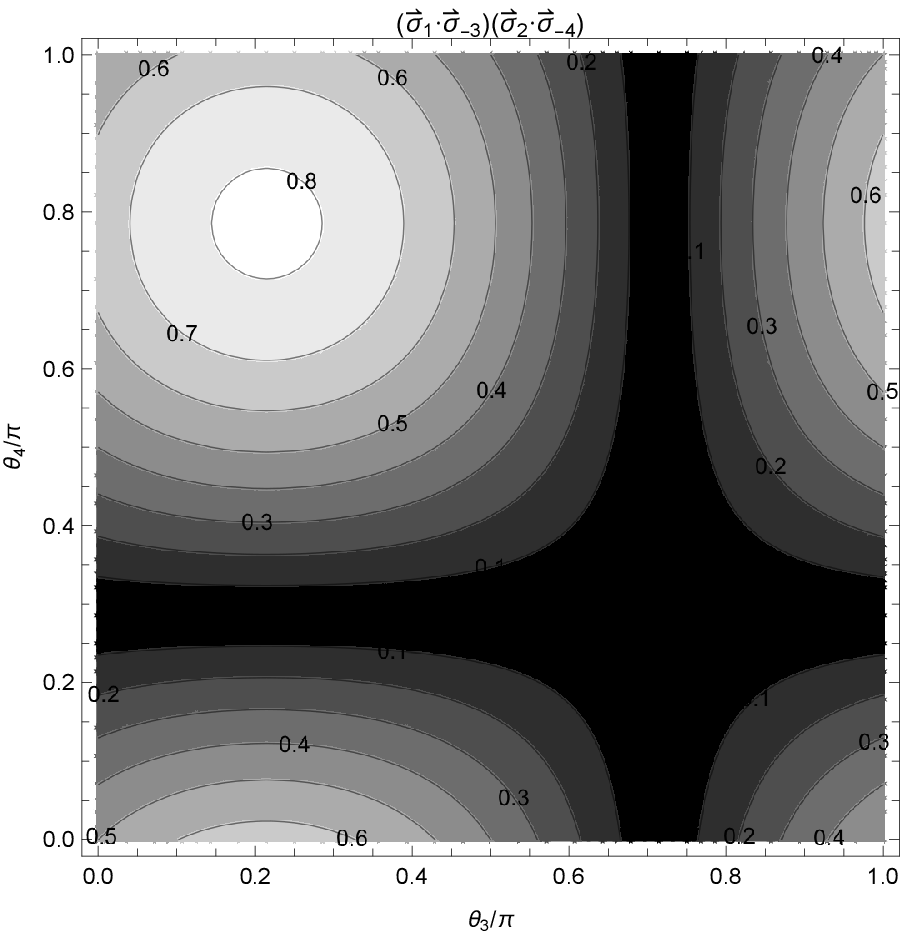}
	\caption{Values of $\hat\sigma_1\cdot\hat\sigma_{-3}$ and  $\hat\sigma_2\cdot\hat\sigma_{-4}$ for $\cos\alpha = 0.8$, and values of $(\hat\sigma_1\cdot\hat\sigma_{-3})(\hat\sigma_2\cdot\hat\sigma_{-4})$ at $\theta_1=\pi/4$ for $\cos\alpha = 0.8$. } 
	\label{fig:fs09}
\end{figure}

To find the probability of emission polarization, we next evaluate the value of $f_{1,-3} + f_{1,-4}$. Let $\theta_i$ be the polarization angle of mode $i$ with respect to the direction $\hat k_1\times \hat k_3$. 
Using the law of cosine, we can write 
\begin{align}
	\hat\sigma_1\cdot\hat\sigma_{-3} &= \cos\theta_1\cos\theta_3 + \sin\theta_1\sin\theta_3\cos\alpha \nonumber \\
	&= \sqrt{\cos^2\theta_1 + \sin^2\theta_1 \cos^2\alpha} \cos(\theta_3-\varphi_3)\nonumber \\
	&= \sqrt{1- \sin^2\theta_1 \sin^2\alpha} \cos(\theta_3-\varphi_3), 
\end{align}
where $\tan\varphi_3 = \cos\alpha \tan\theta_1$.
For $\hat\sigma_2\cdot\hat\sigma_{-4}$, we use $\theta_2 = \theta_1 - \pi/2$, then 
\begin{align}
	\hat\sigma_2\cdot\hat\sigma_{-4} &= \sin\theta_1\cos\theta_4 - \cos\theta_1\sin\theta_4\cos\alpha \nonumber\\
	&= \sqrt{\sin^2\theta_1 + \cos^2\theta_1 \cos^2\alpha} \cos(\theta_4-\varphi_4)\nonumber\\
	&= \sqrt{1 - \cos^2\theta_1 \sin^2\alpha} \cos(\theta_4-\varphi_4), 
\end{align}
where $\tan\varphi_4 = \cos\alpha \tan(\theta_1+\frac\pi2)$. The values of $\hat\sigma_1\cdot\hat\sigma_{-3}$ and  $\hat\sigma_2\cdot\hat\sigma_{-4}$ for different $\alpha$ are plotted in Fig.~\ref{fig:fs09}. For $\alpha\sim 0$, each term reaches its maximum value near $\theta_1$ and $\theta_1-\pi/2$, respectively. The product  $(\hat\sigma_1\cdot\hat\sigma_{-3})(\hat\sigma_2\cdot\hat\sigma_{-4})$ reaches its maximum when $\theta_1 = \frac\pi4$. The same result can be obtained for $f_{1,-4}$. At this angle, the probability that mode 3 is polarized at angle $\theta_3$ and mode 4 is polarized at angle $\theta_4$ is proportional to 
\begin{align}
	f_{1,-3} &+ f_{1,-4} = \left[ \frac{c^2(\bm{k}_1-\bm{k}_3)^2}{(\omega_1-\omega_3)^2-\omega_p^2} -1 \right] \nonumber \\
	& \times (1 + \cos^2\alpha) \cos(\theta_3-\varphi_{30})\cos(\theta_4-\varphi_{40}), \\
	\tan\varphi_{30} &= \cos\alpha, \qquad 
	\tan(\varphi_{40}+\frac\pi2) = \frac{1}{\cos\alpha}.
\end{align} 
This probability function is plotted in Fig.~\ref{fig:fs09}, showing that the two modes have the maximum probability of polarizing at different angles $\tan\varphi_{30}$ and $\tan\varphi_{40}$, respectively.

Therefore, polarization entangled photon pairs can be collected at the azimuthal angle $\alpha$ defined in Eq.~(\ref{eq:alpha}) and the polar angle $\theta_1 = \pi/4$. Now we define the horizontal polarization at the angle $\theta_1$, then the biphoton state can be written as 
\begin{multline}
	|\Psi\rangle =  \frac{1}{\sqrt2} \cos(\varphi_{30} - \varphi_{40}) (|H,V\rangle + |V,H\rangle) \\
	+  \frac{1}{\sqrt2} \sin(\varphi_{30} - \varphi_{40}) (|H,H\rangle + |V,V\rangle). 
\end{multline}
The photon pairs become entangled if either of the two coefficients is significantly larger than the other. Quantitatively, the entanglement can be measured using logarithmic negativity
\begin{equation}
	E_N = \log[2\cos(\varphi_{30} - \varphi_{40})]. 
\end{equation}
Thus, the emitted photon pair is in an entangled state if $\cos(\varphi_{30} - \varphi_{40})>1/2$, which limits the plasma frequency and the accompanying two-pump detuning.

\subsection{Pumps with the same polarization}

We next check the possibility of producing entangled photon pairs using two colinear pump pulses with the same polarization. They produce output fields with the same polarization, which could be different from the pump polarization. 
All the terms, including $f_{1,2}$, $f_{1,-3}$, and $f_{1,-4}$, contribute to the production of photon pairs. Assume the pumps are horizontally polarized and let $\theta_i$ be the polarization angle of mode $i$ with respect to the pump fields. The term $f_{1,2}$ allows creation of both horizontally or vertically polarized photon pairs, \ie 
\begin{multline}
	f_{1,2}	= \left[ \frac{c^2(\bm{k}_1+\bm{k}_2)^2}{(\omega_1+\omega_2)^2-\omega_p^2} -1 \right]  \\
	\times [\cos\theta_3 \cos\theta_4 + \sin\theta_3\sin\theta_4 \cos(2\alpha)]. 
\end{multline}
However, because $\omega_1 + \omega_2 \gg \omega_p$, the term in the first square bracket is strongly suppressed. The photon creation is dominated by the terms $f_{1,-3}$ and $f_{1,-4}$, which equal to $\cos\theta_3\cos\theta_4$ and become $0$ for vertically polarized photon pairs. Therefore, it is impractical to produce polarization entangled photon pairs using two colinear pump pulses with the same polarization.

\subsection{Rate of photon-pair emission}

The photon emission rate is determined by the product of the growth rate $\Gamma_F$ and the pump amplitudes $\alpha_1\alpha_2$. The value of $\Gamma_F$ increases as $\omega_1-\omega_3$ approaches $\omega_p$. But, too near the resonance point, strong spontaneous Raman scattering induces unwanted noise photons. As an example, consider the first pump with a wavelength $\lambda_1=\unit[1]{\mu m}$, and plasma with a density of $\unit[3.5\times10^{15}]{cm^{-3}}$, corresponding to $\omega_p=0.1\omega_2$. The second pump frequency is chosen well beyond the plasma resonance frequency, $\omega_2 = \omega_1-4\omega_p$. In this configuration, we find that the output photon frequency is $\omega_3=\omega_4 = 0.8\omega_1$, or $\lambda_3=\lambda_4=\unit[1.25]{\mu m}$, and they are separated by a $3.76^\circ$ angle. For pump amplitude $\alpha_1\alpha_2=0.01$, the FWM growth rate is $\Gamma_F = 0.0001\omega_1$. Thus a single photon pair ($r\approx 0.15$) could be produced in a $\unit[1.5]{mm}$-long plasma. 

Because the plasma frequency can be flexibly controlled through its density, a wide range of frequencies of the output entangled photon pairs are possible. 
If we instead consider a soft x-ray pump with $\lambda=\unit[10]{nm}$ and $\alpha_1\alpha_2=0.0001$, an entangled photon pair can be created in the same plasma length, if the plasma density is increased to $\unit[3.5\times10^{19}]{cm^{-3}}$.

\subsection{Effects of spontaneous Raman scattering}

In sub-picosecond timescales, spontaneous Raman scattering is the main process that destroys the entanglement. It converts a photon into a different frequency by either absorbing or creating a plasma phonon in the form of Langmuir wave. SRS introduces noise via two routes. First, the thermal phonon scatters one of the entangled photon pairs into a different frequency and a different angle, causing direct loss of quantum correlation. But the scattering is negligible with low photon and phonon numbers. If we assume an equipartition theorem for plasma waves, and consider plasma phonon energy to be  $\hbar\omega_p \sim \unit[0.1]{eV}$ ($\omega_p$ equals to $1/10$ of optical laser frequency), then a cold plasma at $\sim\unit[1]{eV}$ temperature only has an average thermal phonon number of only $\bar{n}_{th}\approx 10$. 

Second, SRS can scatter the pump photons into the output modes. The created photons are not correlated with the entangled photon pairs, reducing the quantum purity of the output states. This process, however, could not be suppressed using techniques for plasma Raman amplifiers, such as a plasma density gradient, because fluctuations of the plasma density could broaden the scattering spectrum and shadow the entangled photon pairs. 
Suppression of SRS requires reducing the thermal photon number of plasma oscillation at frequency $ \omega_1 - \omega_3$. 

\section{Producing quantum squeezed states and Suppressing stimulated Raman scattering }

If multiple photon pairs could be produced, the output becomes a quantum two-mode squeezed state $|\Psi\rangle =  \cosh^{-1}r \sum_{n=0}^\infty e^{in\varphi}\tanh^n r |n,n\rangle$. 
The two output modes have strong quantum correlation similar to the quantum entangled photon pairs but higher brightness than a single photon pair. What is remarkable about two-mode squeezed states is that covariance of their quadratures has below-shot-noise level fluctuations. If the two output modes are linearly combined using a beam splitter, the output becomes a continuous variable (CV) quantum entangled state. The squeezing operation can control the quantum noise which is encoded for quantum communication and manipulated for quantum computation. Because the angle for squeezed quadrature can be continuously tuned, there are unparalleled advantages compared to discrete variable (DV) quantum information, which uses entangled photon pairs. 

To achieve a significant squeezing magnitude, the system needs to have high photon emission rate and low noise input in the squeezed quadrature. 
To fulfill both requirements, we next consider pumping FWM using a bi-color pump with a detuning $\omega_1 - \omega_2 = 2\omega_p$, which yields the maximum FWM interaction rate. This frequency configuration is avoided for producing entangled photon pairs, because it induces rapid scattering from thermal phonons. However, because quantum squeezing reduces noise only in one quadrature, the SRS noise might not degrade the squeezing magnitude if the they are induced in a correlated manner, \ie the noise photon correlation could be made use of for noise cancellation. 

The enhancement of FWM growth rate $\Gamma$ can significantly increase the photon emission rate. 
For example, if we only change the pump wavelength to satisfy $|\omega_{1,2}- \omega_{3,4}|  = \omega_p$ but keep other parameters as given in the last section, $\Gamma$ increases to a value similar to $g\sim 0.015\omega_1$ which is more than one order of magnitude higher than the value given in the last section. Under this condition, A $\unit[1]{mm}$-long plasma and $\unit[30]{fs}$ pump pulse could result in more than $10^9$ photons. If the plasma length and laser duration is doubled, the  photon number would reach $10^{17}$, which contains energy of $\unit[0.01]{J}$.

Wth both FWM and SRS involved, the state evolution is $U = \exp\left[ -\frac{i}{\hbar}\int H_{int}(t) dt \right] = D_3(\beta_3)D_4(\beta_4)S(\xi)$, where $D_i(\beta_i) = \exp(\beta_i \hat{a}_i - \beta_i^* \hat{a}_i^\dag)$ is the displacement operator, $S(\xi) = \exp(\xi \hat{a}_3^\dag \hat{a}_4^\dag -  \xi^* \hat{a}_3\hat{a}_4)$ is the two-mode squeezing operator, $\xi = -i\Gamma_F\beta_1\beta_2$, $\beta_{3,4} = -i(\alpha_{1,2}^*g\bar{p} - \alpha_{2,1}^*g\bar{q})$, and $\bar{p}$ and $\bar{q}$ are the average phonon amplitudes. The output state can then be expressed in the form of a two-mode squeezed coherent state $|\Psi_{out}\rangle = D_3(\beta_3)D_4(\beta_4)S(\xi) |0,0\rangle $. Interestingly, the displacement operators $D_3$ and $D_4$ shift the states with the same quadrature angle and amplitude if $\bar{p}= \bar{q}$. This coherent component might, therefore, be canceled mutually if the modes are linearly combined. 

The bi-color pump configuration has been studied in optical cavities involving oscillating a mechanical oscillator~\cite{Qu_NJP2014}. It shows that asymmetric pump produces strong quantum squeezing and symmetric pump produces phase sensitive amplification. The method of producing squeezing using asymmetric pumps, however, cannot be directly applied to the plasma medium, because it lacks the optical cavity for noise filtering. What could be used here is the combination of symmetric pumping and FWM. While a photon-phonon pair is produced by the high-frequency pump, the phonon is converted into another photon by the low-frequency pump. Thus, it creates correlated photon pairs. It also ensures that the phonon amplitude maintains a finite value, so that the mediating plasma wave can maintain a finite amplitude without an exponential growth. 

However, the thermal noise is driven by plasma wave relaxation which cannot be analyzed through the Hamiltonian itself. Instead, the photon emission process with thermal phonon noise taken into account is described by the quantum Langevin equations (QLE)
\begin{equation}
	\frac{\partial}{\partial t} \bm{\Pi} = \mathcal{L}\bm{\Pi} + \sqrt{2\kappa}\bm{\Pi}_{in}, 
\end{equation}
where we neglected the convection operator, $\bm{\Pi} = (\hat{a}_3, \hat{a}_4^\dag, \hat{p}^\dag, \hat{q})^T$, $\bm{\Pi}_{in} = (0, 0, \hat{p}_{in}^\dag, \hat{q}_{in})^T$, and
\begin{equation}
	\mathcal{L} = \begin{pmatrix}
		0 & i\alpha_1\alpha_2\Gamma_F & -i\alpha_1g & -i\alpha_2g \\
		-i\alpha_1^*\alpha_2^*\Gamma_F & 0 & i\alpha_2^*g & i\alpha_1^*g \\
		i\alpha_1^*g & i\alpha_2 g & -\kappa & 0 \\
		-i\alpha_2^*g & -i\alpha_1 g & 0 &-\kappa
	\end{pmatrix}. 
\end{equation}
Here, $\kappa$ is the plasma wave relaxation rate mainly due to Landau damping at finite plasma temperature, and 
$p_{in}(t)$ and $q_{in}(t)$ are the phonon noise operator associated with the plasma waves. We assume their correlation functions have the similar form with those for Brownian motions
\begin{equation}\label{eq:corr}
	\begin{aligned}
		\langle p_{in}^\dag(t) p_{in}(t') \rangle &=  \bar{n}_p \delta(t-t'), \\ 
		\langle p_{in}(t) p_{in}^\dag(t') \rangle &= (\bar{n}_p+1) \delta(t-t'), \\
		\langle q_{in}^\dag(t) q_{in}(t') \rangle &=  \bar{n}_q \delta(t-t'), \\ 
		\langle q_{in}(t) q_{in}^\dag(t') \rangle &= (\bar{n}_q+1) \delta(t-t'), 
	\end{aligned}	
\end{equation}
where $\bar{n}_{p,q}$ is the average thermal phonon number for mode $p$ and $q$.

The thermal phonon noise is driven by plasma wave relaxations, including Landau damping, collisional damping, and other decoherence processes. These relaxation processes are different from other systems like atoms and mechanical oscillators because the plasma electrons are not directly coupled to external reservoirs. Plasma waves decay mainly from phase mixing due to Landau damping or collision, \ie by coupling to other plasmon modes. It also means that the thermal reservoir is completely determined by the initial condition of the plasma system, in absence of laser interactions, which potentially leads to fewer thermally excited phonons compared to, \textit{e.g.} a mechanical oscillating mirror.

To satisfy $\bar{p}= \bar{q}$, we assume equal pump amplitude with a real value $\alpha_1= \alpha_2 = \alpha$. Then, the QLEs have a simple solution
\begin{align}
	&\hat{a}_3(t) = \cosh(\Gamma_F\alpha^2  t) \hat{a}_3(0) + i \sinh(\Gamma_F\alpha^2  t) \hat{a}_4^\dag(0) \nonumber \\
	& + \int_0^t   \Bigg[ \frac{1+i}{2} \left(\frac{i e^{-\Gamma_F\alpha^2 t'}}{\Gamma_F\alpha^2-\kappa} - \frac{e^{\Gamma_F\alpha^2 t'}}{\Gamma_F\alpha^2+\kappa} \right)   \nonumber \\
	&  + \frac{\Gamma_F\alpha^2 - i\kappa}{(\Gamma_F\alpha^2)^2 - \kappa^2} e^{-\kappa t'} \Bigg] \alpha   g  \sqrt{2\kappa} [p_{in}^\dag(t') + q_{in}(t')]  dt', \\
	&\hat{a}_4(t) = i \sinh(\Gamma_F\alpha^2  t) \hat{a}_3^\dag(0) + \cosh(\Gamma_F\alpha^2  t) \hat{a}_4(0) \nonumber \\
	& \int_0^t   \Bigg[ \frac{1+i}{2} \left(\frac{i e^{-\Gamma_F\alpha^2 t'}}{\Gamma_F\alpha^2-\kappa} - \frac{e^{\Gamma_F\alpha^2 t'}}{\Gamma_F\alpha^2+\kappa} \right)   \nonumber \\
	&  + \frac{\Gamma_F\alpha^2 - i\kappa}{(\Gamma_F\alpha^2)^2 - \kappa^2} e^{-\kappa t'} \Bigg] \alpha   g  \sqrt{2\kappa}  [p_{in}(t') + q_{in}^\dag(t')]  dt'. 
\end{align}
The right hand side of each expression has three terms. The first two terms denote the effect of FWM on the vacuum noise. In the square bracket of the integrand, the first term represents the noise amplification (and deamplification) due to SRS of the high-frequency and low-frequency pumps, and the last term describes the noise due to thermalization. 

\begin{figure}[thp]
	\centering
	\includegraphics[width=0.85\linewidth]{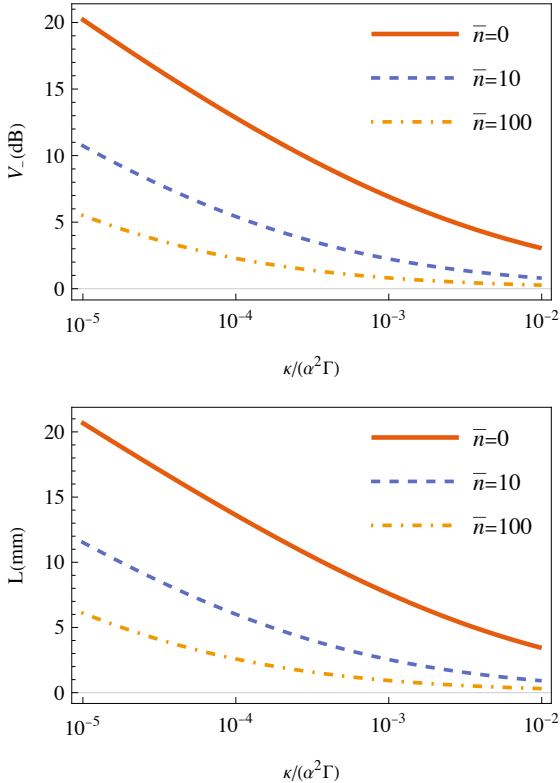}
	\caption{The maximum squeezing magnitude (top) and the optimal plasma length $L=ct_\mathrm{cr}$ (bottom) for different plasma wave damping rate $\kappa$ and thermal phonon number $\bar{n}$. Other parameters are given in the text.} 
	\label{fig:squ}
\end{figure}

The similar forms of $a_3$ and $a_4^\dag$ suggest the possibility of partial cancellation of the noise from phonons. Indeed, the correlation between the two modes is revealed in the $X_3$ and $Y_4$ quadratures, where $X_i = (\hat{a}_i+\hat{a}_i^\dag)/\sqrt2$ and $Y_i = (\hat{a}_i-\hat{a}_i^\dag)/(\sqrt2 i)$ for $i=3$, $4$. Specifically, the correlation function
\begin{align} \label{eq:vm}
	V_- &= \langle (X_3 - Y_4)^2 \rangle = e^{-2\Gamma_F\alpha^2  t} \nonumber \\
	& + \Bigg[ \frac{1-e^{-2\kappa t}}{2\kappa} +  \frac{1-e^{-2\Gamma_F\alpha^2  t}}{2\Gamma_F\alpha^2} - \frac{2(1-e^{-(\Gamma_F\alpha^2 +\kappa)t})}{\Gamma_F\alpha^2 +\kappa} \Bigg] \nonumber \\
	& \times\frac{\alpha^2g^24\kappa}{(\alpha^2\Gamma_F - \kappa)^2}(2\bar{n}_p+1)
\end{align}
only includes terms proportional to $e^{-2\Gamma_F\alpha^2  t}$ and $e^{-2\kappa t}$, but does not have any exponentially growing terms. The first term, representing squeezed vacuum noise, monotonically decreases. The second term, representing thermal phonon noise, saturates at $2\alpha^2g^2(2\bar{n}+1)/[\Gamma\alpha^2( \Gamma\alpha^2+\kappa)] (\gg 1) $. However, thermalization can be negligible in a short time $\kappa t \ll 1$ owing to the small Landau damping rate in cold plasmas. Hence, $V_-\approx 2(g/\alpha\Gamma)^2(2\bar{n}+1)(1-e^{-2\kappa t})$ could be lower than the vacuum level at a certain time. Exactly, the maximum squeezing magnitude is obtained at $t_\mathrm{cr} =  (\alpha^2\Gamma - \kappa)^{-1}  \ln[1+\sqrt\frac{\alpha^2\Gamma(\alpha^2\Gamma-\kappa)^2}{2\kappa\alpha^2 g^2(2\bar{n}+1)}]$. The maximum squeezing magnitude and the optimal plasma length are plotted in Fig.~\ref{fig:squ}. It is seen that the squeezing magnitude crucially relies on a large FWM growth rate $\alpha^2\Gamma$, a small plasma decay rate $\kappa$ and a small thermal phonon number $\bar{n}$. 

The orthogonal correlation exhibits amplified thermal noise due to SRS 
\begin{align}
	V_+ &= \langle (X_3 + Y_4)^2 \rangle = e^{2\Gamma_F\alpha^2  t} \nonumber\\
	&+ \Bigg[ \frac{1-e^{-2\kappa t}}{2\kappa} - \frac{1-e^{2\Gamma_F\alpha^2  t}}{2\Gamma_F\alpha^2  t} - \frac{2(1-e^{(\Gamma_F\alpha^2 -\kappa)t})}{\Gamma_F\alpha^2 - \kappa} \Bigg] \nonumber \\
	&  \times\frac{\alpha^2g^2 4\kappa}{(\alpha^2\Gamma_F + \kappa)^2}(2\bar{n}_p+1). 
\end{align}

\subsection{Single-mode squeezed state}

The two-mode squeezed output can be converted into a single-mode squeezed state using a beam splitter. The output state $\hat c$ can be written as $\hat c=(\hat{a}_3-i\hat{a}_4)/\sqrt2$, which obeys the commutation relation $[\hat c(t), \hat c^\dag(t')] = \delta(t-t')$. The spectra of its quadratures $X_c =(\hat c+\hat c^\dag)/\sqrt2$ and $Y_c = (\hat c-\hat c^\dag)/(\sqrt2 i)$ can be found
\begin{align}
	S_{Xc} &= \langle X_c^2 \rangle = \frac12 e^{-2\Gamma_F\alpha^2  t} \nonumber \\
	& + \Bigg[ \frac{1-e^{-2\kappa t}}{2\kappa} +  \frac{1-e^{-2\Gamma_F\alpha^2  t}}{2\Gamma_F\alpha^2} - \frac{2(1-e^{-(\Gamma_F\alpha^2 +\kappa)t})}{\Gamma_F\alpha^2 +\kappa} \Bigg] \nonumber \\
	& \times\frac{\alpha^2g^2 2\kappa}{(\alpha^2\Gamma_F - \kappa)^2}(2\bar{n}_p+1), \\
	S_{Yc} &= \langle Y_c^2 \rangle = \frac12 e^{2\Gamma_F\alpha^2  t}  \nonumber\\
	&\Bigg[ \frac{1-e^{-2\kappa t}}{2\kappa} - \frac{1-e^{2\Gamma_F\alpha^2  t}}{2\Gamma_F\alpha^2  t} - \frac{2(1-e^{(\Gamma_F\alpha^2 -\kappa)t})}{\Gamma_F\alpha^2 - \kappa} \Bigg] \nonumber \\
	&  \times\frac{\alpha^2g^2 2\kappa}{(\alpha^2\Gamma_F + \kappa)^2}(2\bar{n}_p+1). 
\end{align}
It is seen that the spectrum of $X$ quadrature is squeezed for certain time $t$. The spectrum of $Y$ quadrature, however, shows anti-squeezed noise fluctuation. 

Although squeezing is obtained, excessive noise degrades the state purity $\mathrm{Tr}\rho^2 = 1/\sqrt{\det\sigma}$. It can be obtained using the covariance matrix $\sigma = \linebreak \begin{pmatrix}
	2\langle X_c^2 \rangle & \langle X_cY_c + Y_cX_c \rangle \\
	\langle X_cY_c + Y_cX_c \rangle & 2\langle Y_c^2 \rangle 
\end{pmatrix}$, where $	\langle X_cY_c \rangle = \langle Y_cX_c \rangle = 0$ and we used $\langle X_c \rangle = \langle Y_c \rangle = 0$. Thus, the state purity is $\mathrm{Tr}\rho^2 = 4S_{Xc}S_{Yc}$, which decreases at higher thermal phonon number $\bar{n}_p$ and larger plasma wave relaxation rate $\kappa$.

\section{Conclusions}

In conclusion, we investigated the use of ionized plasmas and ultra-intense laser pulses to generate quantum states of light with high photon flux and broad bandwidth. Our model demonstrates the ability of the relativistic FWM nonlinear susceptibility to convert two pump photons into two output photons at distinct frequencies and angles. The all-optical parametric nature of FWM ensures that the properties of the output photon pairs are solely determined by the pump photon pairs, independent of plasma resonances, though plasma density influences the photon emission rate.
To mitigate classical noise from SRS, the pump frequencies are tailored to ensure substantial detuning of the output frequencies from the Stokes and anti-Stokes sidebands. Employing orthogonally polarized dual-color pumps then enables the generation of polarization-entangled photon pairs.

Setting the pump detuning to twice the plasma frequency enhances both FWM and SRS interaction rates. While this increases SRS noise in both output modes, the quantum correlation of this noise allows for its suppression in one of the quadratures. This configuration facilitates the production of quantum two-mode squeezed states.

It should be noted that the result is obtained by assuming a Brownian-noise-like correlation function [Eqs.~(\ref{eq:corr})] for the plasma thermal phonons. Plasmas follow the same fluctuation-dissipation theorem of statistical dynamics near equilibrium~\cite{Diamond_Itoh_Itoh_2010} that describes the emission of Brownian particles. 
However, note several distinctive features of plasma fluctuations. First, while the Brownian motion of particles is captured in the frequency spectrum of the particle energy distributions, the fluctuation of plasma waves is dependent on both frequency and wavevector. This leads to the second difference between particle fluctuations and wave fluctuations in that the plasma wave relaxation is caused by dephasing of different wavevectors, known as Landau damping, but Brownian motion decays mainly due to Stokes dragging.  Third, plasma couples to electromagnetic waves via its discrete particle distributions, but Brownian particles radiate due to random acceleration by thermal fluctuations.  Nevertheless, the two different models should be equivalent in the limit that the plasma wavelength is shorter than the Debye length and the radiating elements are purely collective plasma wave phonons.

The calculation also neglects the large mismatch of the group velocities between the laser pulses and the plasma waves. A more accurate analysis needs to conduct quantum numerical simulations of the laser-plasma interaction processes to fully characterize the produced quantum states. The simulations should  include more modes that we have to neglect in analytical calculations. The most important modes are the series of Stokes and antiStokes sidebands of the strong squeezed mode. As the output amplitude grows, SRS is shown to be capable of broadening the spectrum into a frequency comb~\cite{Qu_freqcomb} with spacing of $\omega_p$. The comb modes, which are produced through cascaded SRS, could show unique quantum features. These processes, as well as other plasma instabilities, will be investigated using particle-in-cell simulations in future works. 

The advantage of utilizing plasmas lies in their capacity for high photon flux and broad bandwidth generation. Plasma excels in mediating short-wavelength light, like x rays, compared to conventional materials. For example, consider a soft x-ray pump with $\lambda=\unit[10]{nm}$. Even if the pumps have an amplitude of $\alpha_1 \approx \alpha_2=0.00001$, the same output photon numbers could be produced in the same plasma length if the plasma density is increased to $\unit[3.5\times10^{19}]{cm^{-3}}$. With mildly relativistic laser intensities, each millimeter-long plasma can produce nine orders of magnitude growth of photon fluxes before the pump is depleted.

The plasma-based methods for generating ultra-strong and broadband quantum light open new avenues for diverse applications in science and technology, including enhanced X-ray imaging,  quantum metrology, and X-ray nuclear spectroscopy~\cite{coherentXray_np2021, coherentXray_prl2014, coherentXray_PRL2009, coherentXray_s2010, coherentXray_prl2006}. 
A notable example is its potential application in quantum lithography~\cite{lithography_2000, lithography_2001, lithography_2001_2, lithography_GSA, lithography_2012, lithography_2006}, a concept proposed to surpass the Rayleigh diffraction limit using quantum light sources. Currently limited by the low photon count of existing sources, the high photon flux and short wavelengths achievable with plasma-based methods could revolutionize high-resolution photolithography, potentially impacting the multi-billion dollar semiconductor industry significantly.

\begin{acknowledgments}
	This work was supported by NSF Grant No. PHY-2206691. 
\end{acknowledgments}  

\bibliography{entangle}

\end{document}